\documentstyle[titlepage,12pt]{article}
\def\thefootnote{\fnsymbol{footnote}}
\def\@makefnmark{\mbox{$^\@thefnmark$}}
\def\thefootnote{\mbox{\noindent$\fnsymbol{footnote}$}}
\long\def\@makefntext#1{\noindent$^{\@thefnmark}$#1}

\newcommand{\preprint}[2]{\large
\begin{center}
\hspace*{8cm} Preprint EFUAZ\\
\hspace*{8cm} FT-94-{#1}\\
\hspace*{8cm}  {#2} 1994
\end{center}
}

\begin{document}

\title{\vspace*{-3cm} \preprint{04-REV}{November} \vspace*{2cm}
{\Large {\bf More about the $j=0$ relativistic oscillator}}\thanks
{Talk presented at VIII Reuni\'on Anual de Divisi\'on de Part\'{\i}culas
y Campos, Socied\'ad Mexicana de F\'{\i}sica, M\'exico, 15-17 de junio de
1994.}\, \thanks{Submitted to ``Europhys. Letters".}
}

\author{
{\bf Valeri V. Dvoeglazov}\thanks{On leave of absence from
{\it Dept. Theor. \& Nucl. Phys., Saratov State University,
Astrakhanskaya str., 83, Saratov\, RUSSIA}}\,
\thanks{Email: valeri@bufa.reduaz.mx,
dvoeglazov@main1.jinr.dubna.su}\\
Escuela de F\'{\i}sica, Universidad Aut\'onoma de Zacatecas\\
Antonio Doval\'{\i} Jaime\, s/n, Zacatecas 98000, ZAC., M\'exico\\
}

\date{\empty}

\begin{abstract}

I start from the Bargmann-Wigner equations and introduce an interaction
in the form which is similar to a $j=1/2$ case [M. Moshinsky \& A.
Szczepaniak, {\it J. Phys. \,A}{\bf 22}  (1989) L817]. By means of the
expansion of the wave function in the complete set of $\gamma$-matrices
one can obtain  the equations for a system  which could be named as the
$j=0$ Kemmer-Dirac oscillator.

The equations for the components $\phi_1$ and $\phi_2$ are
different from the ones obtained by Y. Nedjadi \& R. Barrett
for the $j=0$ Duffin-Kemmer-Petiau (DKP) oscillator
[{\it J. Phys. \,A}{\bf 27} (1994) 4301].
This fact leads to the dissimilar energy spectrum of
the $j=0$ relativistic oscillator.
\end{abstract}

\maketitle

\renewcommand{\thefootnote}{\arabic{footnote}}

\newpage

Meantime the problem of interaction of a spinor particle with external
fields is understood well, one can not say that on interactions
of bosons and higher spin fermions~\cite{spin}. In the present article I
consider the oscillatorlike interaction of a $j=0$ relativistic particle
in the formalism firstly introduced by Kemmer~\cite{st}-\cite{b-st}. The
problem is shown to be exactly solvable.

A general system of relativistic wave equations for an arbitrary spin
was firstly written by Dirac~\cite{Dirac} and Fierz~\cite{Fiertz}.
In my presentation I use a reformulation of their formalism by
Bargmann and Wigner~\cite{BW}. For the cases of spin-0 and spin-1
the Bargmann-Wigner set\footnote{This name is usually referred
to the case of a symmetric wave function, $j=1$ and higher.
However, it is easy to show that
these equations (\ref{eq:BW1}) describe a  $j=0$ particle in the case of
an antisymmetric wave function, see below.}
reduces to the two equations  which can be written in the form
(e.~g., ref.~\cite{Lurie})
\begin{eqnarray}\label{eq:BW1}
\cases{\left [ i\gamma^\mu \partial_\mu -m \right ] \Psi (x) = 0  &\cr
&\cr
\Psi (x) \left [ i (\gamma^\mu)^T \partial_\mu
-m \right ] = 0,&}
\end{eqnarray}
where the wave function presents
oneself a $4\times 4$ matrix (symmetric in a $j=1$ case and antisymmetric
in a $j=0$ case) and the derivative acts to the left hand in the second
equation.

Let introduce the interaction like a  $j=1/2$ case, ref.~\cite{Mosh},
\begin{eqnarray}
\left [ i\gamma^\mu \partial_\mu - k\gamma^i \gamma^0 r^i -m\right ] \Psi
(x) &=& 0,\\
\Psi (x) \left [ i (\gamma^\mu)^T \partial_\mu - k (\gamma^i \gamma^0)^T
r^i -m \right ] &=& 0.
\end{eqnarray}

Then, let  expand $\Psi$ in terms of a set of 16 $\gamma$- matrices.
It is possible to assure ourselves that the symmetry of the wave function
is preserved in the expansion if the set is chosen as $C$, $\gamma^5 C$,
$\gamma^5 \gamma^\mu C$, providing the antisymmetric part, and $\gamma^\mu
C$, $\sigma^{\mu\nu} C$,  for the symmetric part (this  form of the
interaction does not mix the $j=0$ and $j=1$ states). $C$ is the matrix of
a charge conjugation.

By using the properties: $C (\gamma^\mu)^T C^{-1}= - \gamma^\mu $ and
$C (\sigma^{\mu\nu})^T C^{-1}= -\sigma^{\mu\nu}$ in the case of the spin-0
wave function\footnote{The case of the spin $j=1$ will be reported elsewhere.}
\begin{equation}
\Psi_{[ \alpha\beta ]}= C_{\alpha\beta}\varphi +\gamma^5_{\alpha\tau}
C_{\tau\beta}\tilde \varphi + \gamma^5_{\alpha\delta}
\gamma^\mu_{\delta\tau} C_{\tau\beta} \tilde A_\mu ,
\end{equation}
one can come to
\begin{eqnarray}\label{eq:1}
\cases{m \varphi = 0 &\cr
m \tilde \varphi = -i (\partial_\nu \tilde A^\nu )& \cr
m \tilde A_\mu = -i \partial_\mu \tilde \varphi + k \left [g^{0\nu}
g_{\mu i} +g^{i\nu} g_{\mu 0}\right ] r^i \tilde A_\nu . & }
\end{eqnarray}
Thus, the initial reducible representation
is decomposed into the $(1/2,1/2)$
vector representation, the $(0,0)$ scalar representation and
the trivial (pseudo)scalar representation, similar to
the Duffin-Kemmer-Petiau  algebra.

Without interaction ($k=0$) the above equations coincide (within
the definition of $\kappa$, the constant which is proportional to  mass)
with Eqs. (26.12) in~\cite{b-st1} and
Eqs. ($247,247^\prime$) in ref.~[5a],  what characterizes
the formalism of Kemmer\footnote{This formulation is also contained in
the more general formulation of Dirac~\cite{Dirac} as mentioned in
ref.~\cite{b-st1}. Therefore, I take a liberty to name the equations
(\ref{eq:eq},\ref{eq:eqq}) as the Kemmer-Dirac oscillator.}:
\begin{eqnarray}
\cases{ m\tilde\varphi = -i\partial_\mu \tilde{A}^\mu
&\cr m\tilde {A}_\mu = -i\partial_\mu \tilde \varphi . &}
\end{eqnarray}
After a  substitution of the second equation to the first one they yield
the Klein-Gordon equation for a spinless particle.

For the stationary states $\tilde \varphi (x, t) = \tilde
\varphi (x) \exp (-iEt)$,
$\tilde A_\mu (x, t) = \tilde A_\mu (x) \exp (-iEt) $
the above set (\ref{eq:1}) is rewritten to\footnote{I chose
a dependence of the wave function on time similar to
refs.~\cite{Mosh,Ned}.  If use $\tilde \varphi (x, t) = \tilde\varphi (x)
\exp (iEt)$, $\tilde A_\mu (x, t) = \tilde A_\mu (x) \exp (iEt)$ the
components $\phi_1$ and $\phi_2$ are only interchanged each other and
$\omega \rightarrow -\omega$ in Eqs.  (\ref{eq:eq},\ref{eq:eqq}), what,
surprisingly, does not lead to any change of the energy spectrum.}
\begin{eqnarray}
\cases{m \tilde \varphi =
- E \tilde{A}_{0} -i\nabla \vec{\tilde{A}} & \cr m\tilde{A}_{0} = -E
\tilde\varphi + k (\vec r \vec{\tilde{A}}) & \cr m\vec{\tilde{A}} =
i\nabla \tilde\varphi + k\vec r \tilde{A}_{0}&}
\end{eqnarray}
(cf. with Eqs. (9) in  ref.~\cite{Ned}). I shall
demonstrate that the above
equations describe oscillatorlike system.  The $j=0$ relativistic
oscillators are also considered in~\cite{Deb}-\cite{Dv}.

After simple algebraic transformations  one can  come to
the following set of equations:
\begin{eqnarray}
\cases{(E-m) \phi_1 = \vec p^{\, -} \vec{\tilde{A}} &\cr
(E+m) \phi_2 = \vec p^{\, +} \vec{\tilde{A}} &\cr
\vec p^{\, +} \phi_1 - \vec p^{\, -} \phi_2 = m\vec{\tilde{A}},&}
\end{eqnarray}
where $\vec p ={1\over i} \vec \nabla$,
$\vec p^{\, \pm} = {1\over \sqrt{2}} (\vec p \pm k \vec r)$
and
\begin{equation}
\phi_1 = \frac{\tilde A_0 -\tilde\varphi}{\sqrt{2}},\hspace*{1cm}
\phi_2 = \frac{\tilde A_0 +\tilde \varphi}{\sqrt{2}}.
\end{equation}
Multiplying the first and the second equations by $m$ one  finds
\begin{eqnarray}
\cases{
m(E-m)\phi_1 = \vec p^{\,-} \vec p^{\,+}\phi_1 -
\vec p^{\,-}\vec p^{\,-}\phi_2  &\cr
&\cr
m(E+m)\phi_2 = \vec p^{\,+}\vec p^{\,+} \phi_1 -
\vec p^{\,+}\vec p^{\,-}\phi_2 ,&}
\end{eqnarray}
and acting $m(E+m)$ at the first equation and
$m(E-m)$ at the second one yields
\begin{eqnarray}
\cases{
m^2 (E^2-m^2) \phi_1 = m(E+m)\vec p^{\,-} \vec p^{\,+}\phi_1
 - (\vec p^{\,-}\vec p^{\,-})(\vec p^{\,+}\vec p^{\,+}) \phi_1
 + (\vec p^{\,-}\vec p^{\,-})(\vec p^{\,+} \vec p^{\,-}) \phi_2 &\cr
&\cr
m^2 (E^2 -m^2) \phi_2 = -m (E-m) \vec p^{\,+}\vec p^{\,-} \phi_2
- (\vec p^{\,+} \vec p^{\,+})(\vec p^{\,-} \vec p^{\, -}) \phi_2
+ (\vec p^{\,+} \vec p^{\,+}) (\vec p^{\,-} \vec p^{\,+})\phi_1 .&}
\end{eqnarray}
Finally, by means of the use of the following commutation relations:
\begin{eqnarray}
&&\left [ p_i^+ \, p_j^- \right ]_{-} = ik\delta_{ij}, \qquad
\left [ p_i^{\pm}\,p_j^{\pm} \right ]_{-} = 0, \\
&& \left \{ p_i^- p_j^+ - p_i^+ p_j^- \right \} f (\vec r) = \left [ -ik
\delta_{ij} +k\epsilon_{jik} \hat L_k \right ] f (\vec r), \\
&& \left \{ \vec p^{\,-}\vec p^{\,+} +\vec p^{\,+} \vec p^{\,-}\right \}
f(\vec r) = \left [ \vec p^{\,2} -k^2 \vec r^{\,2}\right ] f(\vec r) ,
\end{eqnarray}
(with $L_k$ being the operator of angular momentum and $k=im\omega$)
for the $j=0$ case  we obtain
\begin{eqnarray}\label{eq:eq}
(E^2 -m^2) \phi_1 &=& \left [ \vec p^{\,2} +m^2 \omega^2 \vec r^{\,2}
+ (E+2m) \omega +\omega^2 \hat L^2\right ] \phi_1 , \\
\label{eq:eqq} (E^2 -m^2) \phi_2 &=& \left [ \vec p^{\,2} +m^2 \omega^2 \vec
r^{\,2}
+ (E-2m) \omega +\omega^2 \hat L^2\right ] \phi_2 .
\end{eqnarray}

In fact,  one has the oscillator-behaved term
($m^2 \omega^2 \vec r^{\,2}$);
however, there are additional terms comparing with Eq. (10)
of the paper~\cite{Ned},  the Duffin-Kemmer-Petiau oscillator.
The operator of the angular momentum $\hat L^2$ does not present
in the equations of ref.~\cite{Ned} and there is no a
dependence of the ``constant"  term on
the energy there. The presence of this term could lead to some
speculations since  one can show  that a consequence of this fact
is the ``splitting" of energy levels in the both of equations.
Namely, one has two roots in each of equations.
Moreover, if pass to the nonrelativistic limit
($E=\epsilon +m c^2$, $\epsilon \ll mc^2$) in Eq. (15)
one has the quantity $(2mc^2 -\hbar\omega)\epsilon$, which could be
equal to zero or even negative. Meantime, the sum of
the remained terms on the {\it rhs} in the first equation
(\ref{eq:eq}) is positive.
Does this fact signify that the oscillator system surveys not
for all frequency values? More detailed analysis presented
below permits us to answer these questions.

Now let seek to solve  Eq. (\ref{eq:eq}).
For identification purposes, in what
follows it is $(E^2_{N,\ell} - m^2)/2m$ rather than $E_{N,\ell}$ which
I seek since the first form reduces to the usual Schr\"odinger
eigenvalue in the non-relativistic limit.

If use the basis functions similar to  ref.~\cite{Mosh}, then
$\hat L^2\phi_{1,2} = \ell_{1,2} ( \ell_{1,2} + 1 )\phi_{1,2}$
and energy eigenvalues of the equation associated with  Eq. (\ref{eq:eq})
could be found from the algebraic equation
\begin{equation}\label{eq:eq1}
{1\over 2m} ( E^2 - m^2) - (E + 2m){\omega\over 2m}
- \ell_1( \ell_1 + 1 ) {\omega^2\over 2m} = ( N_1 + {3\over 2} )\omega ,
\end{equation}
where the principal quantum number is a non-negative integer.
This equation is quadratic in $E$ and has therefore 2 roots.
The solutions of equation (\ref{eq:eq1}) are:
\begin{equation}
{1\over 2 m} (E^2_{\pm} - m^2) =
(N_1+{5\over 2})\omega
+ \left( \ell_1(\ell_1 +  1) + {1\over 2} \right) {\omega^2\over 2m} \pm
\Delta_1 ,
\end{equation}
where
\begin{equation}
\Delta_1   = {\omega\over 2 }
\left( 1 + (2N_1 + 5) \left({\omega\over m}\right) +
\left(\ell_1 + {1\over 2}\right )^2
\left( {\omega \over m} \right)^2 \right)^{{1\over 2}}.
\end{equation}
This formula has structural similarities with the eigenvalues
found for the DKP oscillator, ref.~\cite{Ned}, {\it i.~e.},
it involves the usual
3-dimensional harmonic oscillator energy, a term proportional
to $\ell(\ell + 1)$ which appears as some kind of rotational
energy and a third energy contribution $\Delta$ which is a
complicated  function of the oscillator frequency, $\ell_1$ and $N_1$.

In the limit where the oscillator frequencies are such that
$\hbar\omega \ll mc^2$,
keeping only the first-order term in $\omega$ in the equations  leads to
\begin{eqnarray}
{1\over 2 m} (E^2_{+} - m^2) &\simeq& \epsilon^+ =
( N_1 + 3 )\omega ,\\
{1\over 2 m} (E^2_{-} - m^2) &\simeq& \epsilon^- =
( N_1 + 2 )\omega .
\end{eqnarray}

I now seek to solve  the second equation (\ref{eq:eqq}).
Using the same procedure as above the two eigenvalues of the
energies are:
\begin{equation}
{1\over 2 m} (E^2_{\pm} - m^2) =
(N_2+{1\over 2})\omega
+ \left( \ell_2(\ell_2 + 1) + {1\over 2} \right) {\omega^2\over 2m} \pm
\Delta_2 ,
\end{equation}
where
\begin{equation}
\Delta_2   = {\omega\over 2 }
\left( 1 + (2N_2 + 1) \left({\omega\over m}\right) +
\left(\ell_2 + {1\over 2}\right)^2 \left( {\omega \over m} \right)^2
\right)^{{1\over 2}} .
\end{equation}
In the limit of low frequencies
\begin{eqnarray}
{1\over 2 m} (E^2_{+} - m^2) &\simeq& \epsilon^+ =
( N_2 + 1 )\omega ,\\
{1\over 2 m} (E^2_{-} - m^2) &\simeq& \epsilon^- =
N_2 \,\omega .
\end{eqnarray}

The condition of a compatibility of the set of equations
(\ref{eq:eq},\ref{eq:eqq})
ensures us that $N_1 = N_2 +2$ and $\ell_2 = \ell_1$.
Therefore, in the relativistic region we have two physical (positive and
negative) values of the energy like to the other formulations of  a
interacting $j=0$ relativistic particle. However, a remarkable feature of
the presented formulation is the double degeneracy (in $N$) of the levels
in the limit $\hbar \omega \ll mc^2$ except for the ground level.
Let us note that such a phenomenon has been discovered
in ref.~\cite{Bruce} ({\it cf.} $\epsilon^{\pm}$
with Eqs. (11a,11b) of the cited work).
However, the reasons of a introduction of the matrix structure
in the Klein-Gordon equation have not been explained there.
Next, I would like to note that the quantity $(E^2_{\pm} -m^2)/2m$
is seen from Eqs. (18,19) or (22,23) to be non-negative even in the
high-frequency limit.

In conclusion, let me mention that  a behavior of a scalar
particle in external fields has been considered in many publications, see,
e.~g., for the references~\cite{Sokolov,Bagrov}.  The recent publications,
ref.~\cite{Kruglov}, concern with a solution of the problem of  finding
the energy spectra of  a scalar particle with a polarizability in the
constant magnetic, electric fields and in the field of the  plane
electromagnetic wave.  However, as we learnt, the model of the $j=0$
oscillator with the intrinsic spin structure has very specific
peculiarities what differs it from the other model used, e.~g.,
in descriptions of $\pi$- and $K$- mesons.

\medskip

I acknowledge valuable discussions with A. del Sol Mesa. The help of
Dr.~Y.~Nedjadi is greatly appreciated.

\end{document}